\documentclass[conference, review=true]{IEEEtran}

\IEEEoverridecommandlockouts
\usepackage[colorlinks,
linkcolor=black,
anchorcolor=black,
citecolor=black
]{hyperref}
\usepackage{cite}
\usepackage{amsmath,amssymb,amsfonts}
\usepackage{algorithmic}
\usepackage{graphicx}

\usepackage{textcomp}
\usepackage{xcolor}
\usepackage{longtable}% for long tables
\usepackage{setspace}
\usepackage{array}
\usepackage[linesnumbered,ruled,vlined]{algorithm2e}
\usepackage{multirow}
\usepackage[normalem]{ulem}
\usepackage{lineno}
\useunder{\uline}{\ul}{}
\usepackage{booktabs}
\usepackage{amsmath}
\usepackage{float}
\usepackage{amssymb}
\usepackage{multirow}
\usepackage{graphicx}
\usepackage{amsfonts}
\usepackage{balance}
\usepackage{enumitem}
\usepackage{float}                           
\usepackage{overpic}  
\usepackage[most]{tcolorbox}
\usepackage{xcolor, colortbl}
\def\BibTeX{{\rm B\kern-.05em{\sc i\kern-.025em b}\kern-.08em
    T\kern-.1667em\lower.7ex\hbox{E}\kern-.125emX}}
\begin{document}

\title{When SparseMoE Meets Noisy Interactions: An Ensemble View on Denoising Recommendation
\thanks{{$^\dag$}Equal contribution.

{*}Corresponding author.

This work was supported in part by grants from the National Science and Technology Major Project (under grant 2021ZD0111802), the National Natural Science Foundation of China (under grant 62406096), the China Postdoctoral Science Foundation (under grant 2024M760722), and the Fundamental Research Funds for the Central Universities 
 (under grant JZ2024HGQB0093).}}
%This work was supported in part by grants from the National Science and Technology Major Project (under grant 2021ZD0111802), the National Natural Science Foundation of China (under grants 62406096), and the Fundamental Research Funds for the Central Universities 
% (under grant JZ2024HGQB0093).

% \author{
% % \IEEEauthorblockN{
% \textit{name, name, name}
% % }
% \\
% \\
% % \IEEEauthorblockN{
% School, University, city, China}
% % }
\author{
% \IEEEauthorblockN{
% \textit{Weipu Chen{$^\dag$}, Zhuangzhuang He{$^\dag$}, Fei Liu{*}}
% % }
% \\
% \\
% % \IEEEauthorblockN{
% School of Computer Science and Information Engineering, Hefei University of Technology, Hefei, China}
% }

% \and
\IEEEauthorblockN{Weipu Chen{$^\dag$}, Zhuangzhuang He{$^\dag$}, Fei Liu{*}}
\IEEEauthorblockA{\textit{School of Computer Science and Information Engineering} \\
\textit{Hefei University of Technology}\\
Hefei, China \\
\{weipuchenn, hyicheng223\}@gmail.com, feiliu@mail.hfut.edu.cn}}
% \and

% }

%
% For example:
% ------------
%\address{School\\
%	Department\\
%	Address}
%
% Two addresses (uncomment and modify for two-address case).
% ----------------------------------------------------------
%\twoauthors
%  {A. Author-one, B. Author-two\sthanks{Thanks to XYZ agency for funding.}}
%	{School A-B\\
%	Department A-B\\
%	Address A-B}
%  {C. Author-three, D. Author-four\sthanks{The fourth author performed the work
%	while at ...}}
%	{School C-D\\
%	Department C-D\\
%	Address C-D}
%

\maketitle

\begin{abstract}
%Learning from implicit feedback in recommendation systems is an important research topic. Implicit feedback is the most common form of feedback in recommendation systems, but it also contains more noise. To address the noise issue, researchers have proposed various methods. However, most of them handle all noisy datasets with a model that has constant parameters and structure, which inevitably leads to inadequacies in model adaptability and generalization performance.  
Learning user preferences from implicit feedback is one of the core challenges in recommendation. The difficulty lies in the potential noise within implicit feedback. Therefore, various denoising recommendation methods have been proposed recently. However, most of them overly rely on the hyperparameter configurations, inevitably leading to inadequacies in model adaptability and generalization performance.  
In this study, we propose a novel Adaptive Ensemble Learning (AEL) for denoising recommendation, which employs a sparse gating network as a brain, selecting suitable experts to synthesize appropriate denoising capacities for different data samples. To address the ensemble learning shortcoming of model complexity and ensure sub-recommender diversity, we also proposed a novel method that stacks components to create sub-recommenders instead of directly constructing them. Extensive experiments across various datasets demonstrate that AEL outperforms others in kinds of popular metrics, even in the presence of substantial and dynamic noise. Our code is available at https://github.com/cpu9xx/AEL.

\end{abstract}


\section{Introduction}
\begin{figure}[t!]
\centerline{\includegraphics[width=0.5\textwidth]{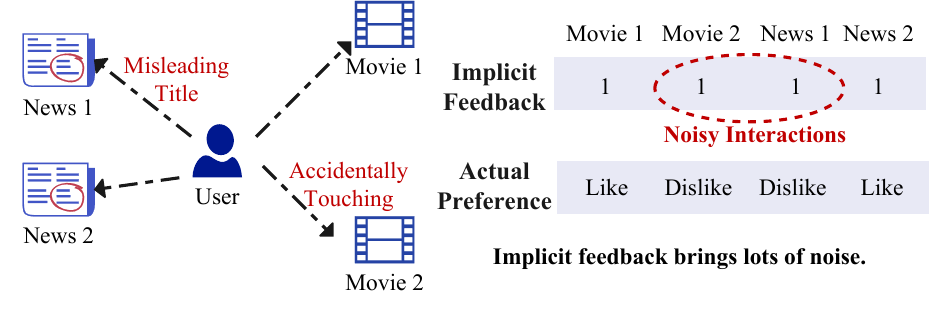}}
\caption{Implicit feedback brings lots of noise in the recommendation task. Denoising recommendation becomes a hotpot.}\label{fig:intro}
\vspace{-15pt}
\end{figure}

% 1. 推荐背景，引入noise
The rise of social media and global communication has led to an explosion of information, making personalized recommendation for users a key challenge. In recent years, learning user preferences from implicit feedback has become a mainstream strategy, as implicit signals like clicks are easier to collect than explicit ratings. However, as shown in Fig.~\ref{fig:intro}, users might accidentally click on items that do not reflect their true preferences~\cite{sun1,sun2,sun3, hui1,bai1,bai2,wang1,wang2}. Recent studies~\cite{DeCA, T-CE, DCF} refer to this type of interaction as \textit{noisy interaction}. Moreover, treating noisy interactions as clean ones can negatively impact recommendation performance~\cite{T-CE, DeCA, post-click}. As a result, denoising recommendation methods~\cite{DeCA, T-CE, DCF, robust-survey, one_class, WRMF, Sampler_Design, knowledge_refine} have emerged as a popular research direction, aiming to learn users' true preferences from implicit feedback.

Most recent efforts can be divided into reweight-based methods~\cite{T-CE, AutoDenoise-weight, BOD} and drop-based methods~\cite{T-CE, AutoDenoise-reinforce, Sampler_Design, missing_data, user_exposure}.
However, these denoising methods still have non-neglectable limitations. 1) Reweight-based methods depend heavily on accurately identifying and assigning appropriate weights to noisy interactions. But, misclassifying numerous noisy interactions can result in suboptimal weight assignments, leading to poor performance. 2) Drop-based methods may drop valuable samples when denoising, leading to a loss of information that could contribute to a better understanding of user preference. Moreover, 
the performance of the both denoising methods overly relies on the hyperparameter configurations, demanding to adjust the denoising capacities case by case to gain notable results. Therefore, how to adaptively adjust the denoising capability without complex hyperparameter tuning also poses a key limitation for denoising recommendation.

To overcome the above limitations, we consider leveraging ensemble learning~\cite{voting,bagging,xgboost}, which is effective in adapting to different application scenarios. We expect this approach routes inputs to specialized sub-recommenders, enabling efficient noise filtering tailored to varying data patterns while minimizing the need for manual tuning.
However, there are two challenges to achieving the above objectives.
\textbf{C1: Increasing model complexity due to multiple sub-recommenders.} Directly creating multiple sub-recommenders increases the model complexity. This makes optimization difficult and raises computational costs, reducing the practicality for large-scale recommendation systems.
\textbf{C2: Overfitting caused by static weight distributions.} General ensemble methods may fix weights for sub-recommenders based on training data causing overfitting. It may perform well on the training set but show poor results on out-of-distribution data, making it less effective in real-world scenarios where data varies.

Therefore, we propose an \textbf{\uline{A}}daptive \textbf{\uline{E}}nsemble \textbf{\uline{L}}earning (AEL) for denoising recommendation. 
Inspired by SparseMoE (Sparse Mixture-of-Experts)~\cite{SMoE}, AEL consists of three experts and a sparse gating network. Specifically, we construct the experts of the denoising module in a novel approach and introduce an adaptive ensemble module to fundamentally address adaptability insufficiency. For \textbf{C1}, we propose a novel method that stacks components to construct sub-recommenders. It not only reduces the model complexity but also ensures sub-recommender diversity. For \textbf{C2}, we creatively introduce a sparse gating network, which can analyze the performance of each sub-recommender and generate suitable dynamic weight distributions for current input.

In summary, our contributions are as follows:
\begin{itemize}[leftmargin=*]
  \item We pioneer an effort to analyze the pitfalls of existing denoising recommendation methods that suffer from low adaptability due to an excessive number of hyperparameters.
  
  \item We propose an Adaptive Ensemble Learning (AEL) method for denoising recommendation. Specifically, we design a novel process for constructing sub-recommenders to address the issue of model complexity. Additionally, we develop an adaptive ensemble module to tackle the overfitting problem caused by static weight distribution.
  
  \item We conduct extensive experiments compared to cutting-edge methods and investigate various modules of our model, demonstrating the effectiveness and adaptability of AEL.
\end{itemize}

\section{Proposed Model}

\begin{figure}[t!]
\centerline{\includegraphics[width=0.5\textwidth]{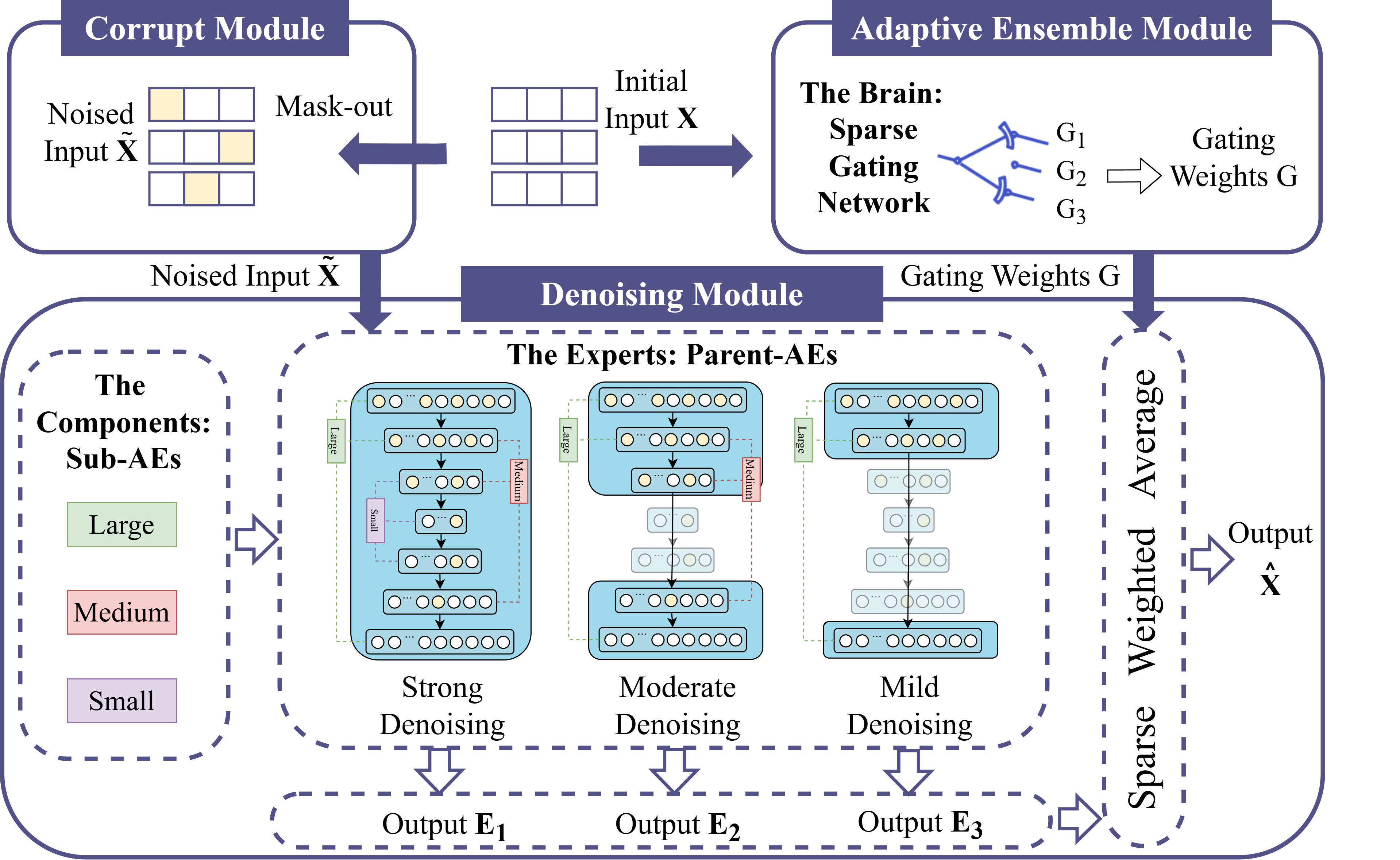}}
\caption{An overview of proposed AEL.}\label{fig:model}
\vspace{-10pt}
\end{figure}

AEL contains three modules: the denoising module, the corrupt module, and the adaptive ensemble module. In the denoising module, we first construct three \textit{sub-Autoencoders} (sub-AEs) as components based on collaborative denoising autoencoder~\cite{CDAE,DAE}. Then, we vary the denoising capacities of three parent-AEs and reduce model size using a novel method, which first creates three sub-AEs as components, then stacks them to construct heterogeneous \textit{parent-Autoencoders} (parent-AEs). We also introduce a corrupt module to improve robustness by partially corrupting initial input, preventing sub-AEs from simply learning the identity function. The adaptive ensemble module achieves denoising capacity adaptability. It contains an improved \textit{sparse gating network}~\cite{SMoE} as a brain, which can analyze the historical performance of parent-AEs, and automatically select the two most suitable parent-AEs to synthesize appropriate denoising capacity for current input data. An overview of AEL is shown in Fig.~\ref{fig:model}. 

%\subsection{Adaptive Ensemble Learning}
\subsection{The Components: Sub-Autoencoders}
To save computing resources and reduce model size, we invent a novel method that allows the experts in SparseMoE to share parameters. We design and form three sub-AEs based on Collaborative Denoising Autoencoder (CADE), named Large, Medium, and Small, as components, then stack them to construct parent-AEs as experts.  
Specifically, each sub-AE consists of two functional modules: the encoder and the decoder. Their main difference is the hidden dimension. In the encoding process, sub-AEs encode additional latent vectors for user \( V_u \) to provide improved recommendation. Large also utilizes the corrupt module to corrupt the input \( x_u \) to \( \tilde{x}_u \) using a mask-out technique. Its encoder and decoder functions can be described by the following formula:
\begin{equation}
 \text{Encoder: } {z}_u^{(1)} = \sigma\left(W_{\text{id}}^T \cdot u + W_{\text{item}}^T \cdot \tilde{x_u} + b_{\text{item}}\right),
\end{equation}
\begin{equation}
\text{Decoder:   } \hat{x}_u = \sigma(W_d^T \cdot z_u^{(1)} + b_d),
\end{equation}
where \( \tilde{x_u} \in \mathbb{R}^{1 \times D} \) is the input, and \( \hat{x_u} \) is the output. \( z_u \in \mathbb{R}^{1 \times K} \) is the latent feature representation of the input \( \tilde{x_u} \). \( K \) is the hidden dimension and \( K \ll D \). \( W_{\text{id}}^T \), \( W_{\text{item}}^T \), and \( W_d^T \) are the weight matrices, \( b_{\text{item}} \) and \( b_d \) are the bias vectors of the network, \( \sigma\left(\cdot\right)\) is the activation function.

The other sub-AEs, Medium and Small, respectively take the hidden vector $z_u^{(j)}$ of the previous sub-AE as input. Details of three sub-AEs, including their encoders, decoders, and loss functions, are shown in Table~\ref{tab:Characteristics of sub-AEs}.
Parameters of sub-AEs are learned by minimizing the average reconstruction error:
\begin{equation}
\arg\min\frac{1}{U}\sum\nolimits_{u=1}\nolimits^{U}\mathbb{E}_{p(\tilde{\boldsymbol{x}}|\boldsymbol{x})}\left[\ell\left(\hat{\boldsymbol{x}},{\boldsymbol{\tilde{x}}}\right)\right]+\mathcal{R}\left(\boldsymbol{W},\boldsymbol{W}^{\prime},\boldsymbol{V},\boldsymbol{b},\boldsymbol{b}^{\prime}\right),
\end{equation}
where $\mathcal{R}\left(\cdot\right)$ is the regression term with the squared \textit{L}$_2$ Norm.
\begin{equation}
\mathcal{R}\left(\cdot\right)=\frac{\lambda}{2}\left(\|\boldsymbol{W}\|_{2}^{2}+\|\boldsymbol{W}'\|_{2}^{2}+\|\boldsymbol{V}\|_{2}^{2}+\|\boldsymbol{b}\|_{2}^{2}+\|\boldsymbol{b}'\|_{2}^{2}\right).
\end{equation}

\begin{table*}[h!]
\centering
\caption{Details of sub-AEs.}
% \vspace{-7pt}
\label{tab:Characteristics of sub-AEs}
\resizebox{0.99\textwidth}{!}{
\begin{tabular}{ccccc}
  \toprule
  \textbf{Component} & \textbf{Hidden Dim.} & \textbf{Encoder} & \textbf{Decoder} & \textbf{Loss Func.} \\
  \midrule
Large  & 128 & $z_u^{(1)} = \sigma\left(W_{\text{id}}^T \cdot u + W_{\text{item}}^T \cdot \tilde{x_u} + b_{\text{item}}\right)$ & $ \hat{x}_u = \sigma(W_d^T \cdot z_u^{(1)} + b_d)$ & $\frac{\sum_{i=1}^{n} (\hat{R}_{ui} - R_{ui})^2}{m}$\\
%\hline
Medium & 48 & $z_u^{(2)} = \sigma\left(W_{\text{id}}^{\prime T} \cdot u + W_{\text{Large}}^T \cdot  z_u^{(1)} + b_{\text{Large}}\right)$ & $\hat{z}_u^{(1)} = \sigma(W_d^{\prime T} \cdot z_u^{(2)} + b_d^{\prime})$ & $\frac{1}{n^{(1)}} \sum_{i=1}^{n^{(1)}} (\hat{z}_{ui}^{(1)} - z_{ui}^{(1)})^2$\\
%\hline
Small & 12 & $z_u^{(3)} = \sigma\left(W_{\text{id}}^{\prime\prime T} \cdot u + W_{\text{Medium}}^T \cdot z_u^{(2)} + b_{\text{Medium}}\right)$ & $\hat{z}_u^{(2)} = \sigma(W_d^{\prime\prime T} \cdot z_u^{(3)} + b_d^{\prime\prime})$ & $\frac{1}{n^{(2)}} \sum_{i=1}^{n^{(2)}} (\hat{z}_{ui}^{(2)} - z_{ui}^{(2)})^2$\\
\bottomrule
\end{tabular}}
\vspace{-0.5cm}
\end{table*}

\subsection{The Experts: Parent-Autoencoders}
We hope sub-AEs, the three autoencoders with varying hidden dimensions, will have different denoising capacities so that we can take them as experts. However, the difference in hidden dimension merely illustrates that they have different numbers of parameters to capture the features. Considering that the features may be represented more compactly or more loosely, a causal link cannot be established between the hidden dimension and the denoising capacity of an autoencoder.

Therefore, we stack sub-AEs to construct three parent-AEs, respectively named Mild Denoising, Moderate Denoising, and Strong Denoising. Since the hidden dimension is smaller than the input vector dimension, a sub-AE cannot fully recover its original input after encoding and decoding. This process performs denoising. Leveraging this characteristic, we ensure that parent-AEs have varying denoising capacities by stacking sub-AEs in different ways.

\subsection{The Brain: Sparse Gating Network}
We construct three parent-AEs as experts using a novel method that significantly reduces the model size, as mentioned above. However, the denoising capacity of an individual parent-AE remains fixed. To achieve adaptability to dynamic noise intensities, we introduce an improved sparse gating network~\cite{SMoE} to manage these parent-AEs. %It will select the two most suitable parent-AEs to synthesize the appropriate denoising capacity for varying scenarios.

The sparse gating network utilizes the \textit{Noisy Top-K Gating} strategy for selecting experts. In this approach, tunable Gaussian noise is added before applying the softmax function. This helps in load balancing and ensures that only suitable experts are activated for each input.
\begin{equation}
G(x)_i=Softmax(KeepTopK(H_i(x),k)),
\label{eq:G(x)}
\end{equation}
\begin{equation}
H(x)_i=(x\cdot W_g)_i+StdNormal()\cdot Softplus((x\cdot W_{n})_i),
\label{eq:H(x)}
\end{equation}
where $W_g$ and $W_{n}$ are trainable weight matrices, $G(x)_i$ is the output of the gating network, and \(KeepTopK\) function retains the top \(k\) values and sets the rest to $-\infty $.

We combine the outputs of the gating network $G(x)_i$ and experts $E_i(x)$ to generate overall predictions:
\begin{equation}
\hat{x}=\sum\nolimits_{i=1}\nolimits^{3}G(x)_iE_i(x).
\end{equation}
Additionally, to ensure that all the parent-AEs have similar workloads and avoid imbalance, we introduced two additional loss functions:  
\begin{equation}
L_{importance}(X)=w_{i}\cdot CV(\sum\nolimits_{x\in X}G(x))^2,
\label{eq:Limportant(x)}
\end{equation}
\begin{equation}
L_{load}(X)=w_{l}\cdot CV(\sum\nolimits_{x\in X}P(x,i))^2,
\label{eq:Lload(x)}
\end{equation}
where $CV\left(\cdot\right)$ is the squared coefficient of variation function, $P(x,i)$ is the probability that $G(x)_i$ is nonzero, $w_{i}$ and $w_{l}$ are hand-tuned scaling factors, $X$ is a batch of $n$ training examples $x_u$.
With these additional loss functions, the overall loss function can be described as follows:
\begin{equation}
L(X) = L_{importance}(X) + L_{load}(X) + \frac{1}{n}\sum\nolimits_{x\in X}(\hat{x_u}-x_u)^2.
\end{equation}

\vspace{-0.5cm}
\subsection{Model Size discussion}
AEL employs ensemble learning to achieve denoising capacity adaptability. However, due to the novel method we used for constructing the parent-AEs, AEL addresses the model complexity issues that other ensemble models have. Table~\ref{tab:parameters statistics} compares the number of parameters across different models, including GMF~\cite{GMF+NeuMF}, NeuMF~\cite{GMF+NeuMF}, and DeCA~\cite{DeCA}.

\vspace{-7pt}
\begin{table}[H]

\begin{center}
\caption{Statistics of model parameters.}
\vspace{-7pt}
\label{tab:parameters statistics}
\begin{tabular}{ccccc}
  \toprule
  \textbf{Model} & \textbf{GMF} & \textbf{NeuMF} & \textbf{DeCA(GMF)} & \textbf{Ours} \\
  \midrule
  Parameters & 7,371,905 & 7,371,937 & 14,743,810 & 8,695,336 \\
  %Baby & 3,557 & 5,788 & 26,688 & 0.99870 \\
  
  \bottomrule
\end{tabular}
\end{center}
\vspace{-0.5cm}
\end{table}

\begin{table}[t]
\begin{center}
\caption{Statistics of datasets.}
\vspace{-6pt}
% \vspace{-7pt}
\label{tab:dataset statistics}
\small
\begin{tabular}{crrrc}
  \toprule
  \textbf{Dataset} & \textbf{\#Users} & \textbf{\#Items} & \textbf{\#Interactions} & \textbf{Sparsity} \\
  \midrule
  MovieLens & 943 & 1,682 & 100,000 & 0.93695 \\
%   \hline
  Modcloth & 44,784 & 1,020 & 99,893 & 0.99781 \\
  Adressa & 212,231 & 6,596 & 419,491 & 0.99970 \\
  %Baby & 3,557 & 5,788 & 26,688 & 0.99870 \\
  \bottomrule
\end{tabular}
\end{center}
\vspace{-20pt}
\end{table}
%You can see in the table that our AEL model contains 8,695,336 parameters, whereas the baseline single models GMF and NeuMF have 7,371,905 and 7,371,937 parameters, respectively. This suggests that the complexity of AEL is quite comparable to that of the baseline models.
\begin{table*}[t!]
% \tiny
\renewcommand\arraystretch{1}
\caption{Performance comparison. R and M denote Recall and MRR, respectively. A significant improvement between \textbf{best-performing} and $\underline{\text{runner-up}}$ is marked with * (i.e., two-sided t-test with $p \textless 0.05$) and ** (i.e., two-sided t-test with $0.05 \leq p \textless 0.1$).}
\vspace{-0.3cm}
\centering
\small
\label{overrall compa}
\vspace{0.25cm}
\resizebox{0.98\textwidth}{!}{
\begin{tabular}{@{}c|llll|llll|llll@{}}
\toprule[1.2pt]
\multicolumn{1}{c}{\textbf{Dataset}} & \multicolumn{4}{|c}{\textbf{Adressa}} & \multicolumn{4}{|c}{\textbf{MovieLens}} & \multicolumn{4}{|c}{\textbf{Modcloth}}   \\ 
\midrule
\multicolumn{1}{c}{\textbf{Metric}}& \multicolumn{1}{|c}{\textbf{R@5}} & \multicolumn{1}{c}{\textbf{R@20}} & \multicolumn{1}{c}{\textbf{M@5}}  & \multicolumn{1}{c}{\textbf{M@20}} & \multicolumn{1}{|c}{\textbf{R@5}} & \multicolumn{1}{c}{\textbf{R@20}} & \multicolumn{1}{c}{\textbf{M@5}} & \multicolumn{1}{c}{\textbf{M@20}} & \multicolumn{1}{|c}{\textbf{R@5}} & \multicolumn{1}{c}{\textbf{R@20}} & \multicolumn{1}{c}{\textbf{M@5}} & \multicolumn{1}{c}{\textbf{M@20}} \\
\midrule
\midrule
\multirow{1}{*}{GMF} & \multicolumn{1}{l}{0.1017} & \multicolumn{1}{l}{0.2111} & \multicolumn{1}{l}{0.0764} & 0.0937 & \multicolumn{1}{l}{0.0567} & \multicolumn{1}{l}{0.1479} & \multicolumn{1}{l}{0.5106} & 0.5293  & \multicolumn{1}{l}{0.0668} & \multicolumn{1}{l}{0.2356} & \multicolumn{1}{l}{0.0453} & 0.0611 \\

\multirow{1}{*}{NeuMF} & \multicolumn{1}{l}{\uline{0.1716}} & \multicolumn{1}{l}{0.3109} & \multicolumn{1}{l}{\uline{0.1453}} & \uline{0.1651} & \multicolumn{1}{l}{0.0901} & \multicolumn{1}{l}{0.2552} & \multicolumn{1}{l}{0.6557} & 0.6692 & \multicolumn{1}{l}{0.0776} & \multicolumn{1}{l}{0.2380} & \multicolumn{1}{l}{0.0529} & 0.0684 \\

\multirow{1}{*}{LightGCN} & \multicolumn{1}{l}{0.0896} & \multicolumn{1}{l}{0.2193} & \multicolumn{1}{l}{0.0753} & 0.0938 & \multicolumn{1}{l}{0.0856} & \multicolumn{1}{l}{0.2388} & \multicolumn{1}{l}{0.6568} &  0.6668 & \multicolumn{1}{l}{0.0631} & \multicolumn{1}{l}{0.2205} & \multicolumn{1}{l}{0.0395} & 0.0547 \\

\multirow{1}{*}{DeCA (CDAE)} & \multicolumn{1}{l}{0.1619} & \multicolumn{1}{l}{\uline{0.3188}} & \multicolumn{1}{l}{0.1320} & 0.1574 & \multicolumn{1}{l}{\uline{0.0958}} & \multicolumn{1}{l}{0.2511} & \multicolumn{1}{l}{\uline{0.6935}} & \uline{0.7040} & \multicolumn{1}{l}{\uline{0.0903}} & \multicolumn{1}{l}{\uline{0.2602}} & \multicolumn{1}{l}{\uline{0.0540}} & \uline{0.0706} \\

\multirow{1}{*}{DeCA (GMF)} & \multicolumn{1}{l}{0.1239} & \multicolumn{1}{l}{0.2144} & \multicolumn{1}{l}{0.0932} & 0.1064 & \multicolumn{1}{l}{0.0704} & \multicolumn{1}{l}{0.1686} & \multicolumn{1}{l}{0.5904} &  0.6064 & \multicolumn{1}{l}{0.0754} & \multicolumn{1}{l}{0.2461} & \multicolumn{1}{l}{0.0525} & 0.0682 \\

\multirow{1}{*}{TCE (GMF)} & \multicolumn{1}{l}{0.1251} & \multicolumn{1}{l}{0.2155} & \multicolumn{1}{l}{0.1010} & 0.1149 & \multicolumn{1}{l}{0.0933} & \multicolumn{1}{l}{\uline{0.2577}} & \multicolumn{1}{l}{0.6786} &  0.6892 & \multicolumn{1}{l}{0.0601} & \multicolumn{1}{l}{0.1743} & \multicolumn{1}{l}{0.0458} & 0.0574 \\

\multirow{1}{*}{RCE (GMF)} & \multicolumn{1}{l}{0.1328} & \multicolumn{1}{l}{0.2145} & \multicolumn{1}{l}{0.1025} & 0.1147 & \multicolumn{1}{l}{0.0886} & \multicolumn{1}{l}{0.2363} & \multicolumn{1}{l}{0.6579} &  0.6684 & \multicolumn{1}{l}{0.0614} & \multicolumn{1}{l}{0.1788} & \multicolumn{1}{l}{0.0478} & 0.0598 \\
%\cmidrule(l){2-13}
\midrule
\multirow{1}{*}{\textbf{Ours}} & \multicolumn{1}{l}{\textbf{0.1749*}} & \multicolumn{1}{l}{\textbf{0.3199**}} & \multicolumn{1}{l}{\textbf{0.1491*}} & \textbf{0.1718*} & \multicolumn{1}{l}{\textbf{0.1087*}} & \multicolumn{1}{l}{\textbf{0.2933*}} & \multicolumn{1}{l}{\textbf{0.7554*}} &  \textbf{0.7648*} & \multicolumn{1}{l}{\textbf{0.0954*}} & \multicolumn{1}{l}{\textbf{0.2753*}} & \multicolumn{1}{l}{\textbf{0.0626*}} & \textbf{0.0800*} \\

\bottomrule[1.2pt]
\end{tabular}}
\vspace{-0.2cm}
\end{table*}

\begin{table*}[ht!]
\renewcommand\arraystretch{1}
\caption{Performance comparison of different \( k \) across three different datasets. R and M denote Recall and MRR, respectively.}
%\vspace{-1em}
\centering
\label{k comparison}
%\vspace{0.25cm}
% \resizebox{0.9\textwidth}{!}{
% \renewcommand{\arraystretch}{1.5} % 调整行距
\footnotesize
\setlength{\tabcolsep}{7.7pt}
\begin{tabular}{@{}c|llll|llll|llll@{}}
\toprule[1pt]
\multicolumn{1}{c}{\textbf{Dataset}} & \multicolumn{4}{|c}{\textbf{Adressa}} & \multicolumn{4}{|c}{\textbf{MovieLens}} & \multicolumn{4}{|c}{\textbf{Modcloth}}   \\ 
\midrule
$k$& \multicolumn{1}{|c}{\textbf{R@5}} & \multicolumn{1}{c}{\textbf{R@20}} & \multicolumn{1}{c}{\textbf{M@5}}  & \multicolumn{1}{c}{\textbf{M@20}} & \multicolumn{1}{|c}{\textbf{R@5}} & \multicolumn{1}{c}{\textbf{R@20}} & \multicolumn{1}{c}{\textbf{M@5}} & \multicolumn{1}{c}{\textbf{M@20}} & \multicolumn{1}{|c}{\textbf{R@5}} & \multicolumn{1}{c}{\textbf{R@20}} & \multicolumn{1}{c}{\textbf{M@5}} & \multicolumn{1}{c}{\textbf{M@20}} \\
\midrule
\midrule
\multirow{1}{*}{3} & \multicolumn{1}{l}{\textbf{0.1759}} & \multicolumn{1}{l}{\uline{0.3192}} & \multicolumn{1}{l}{0.1480} & 0.1704 & \multicolumn{1}{l}{\textbf{0.1087}} & \multicolumn{1}{l}{\uline{0.2903}} & \uline{0.7535} & \uline{0.7629}  & \multicolumn{1}{l}{0.0940} & \multicolumn{1}{l}{0.2717} & \multicolumn{1}{l}{0.0619} & 0.0788 \\

\multirow{1}{*}{2} & \multicolumn{1}{l}{{0.1749}} & \multicolumn{1}{l}{\textbf{0.3199}} & \multicolumn{1}{l}{\textbf{0.1491}} & \textbf{0.1718} & \multicolumn{1}{l}{\textbf{0.1087}} & \multicolumn{1}{l}{\textbf{0.2933}} & \multicolumn{1}{l}{\textbf{0.7554}} & \textbf{0.7648} & \multicolumn{1}{l}{\textbf{0.0954}} & \multicolumn{1}{l}{\textbf{0.2753}} & \multicolumn{1}{l}{\textbf{0.0626}} & \textbf{0.0800} \\

\multirow{1}{*}{1} & \multicolumn{1}{l}{\uline{0.1757}} & \multicolumn{1}{l}{0.3191} & \multicolumn{1}{l}{\uline{0.1488}} & \uline{0.1712} & \multicolumn{1}{l}{\textbf{0.1087}} & \multicolumn{1}{l}{{0.2814}} & \multicolumn{1}{l}{{0.7528}} & {0.7615} & \multicolumn{1}{l}{\uline{0.0947}} & \multicolumn{1}{l}{\uline{0.2734}} & \multicolumn{1}{l}{\uline{0.0623}} & \uline{0.0796} \\

\bottomrule[1pt]
\end{tabular}
\vspace{-0.1cm}
\end{table*}

\section{Experiment}

\subsection{Experimental Settings}
\noindent \textbf{Dataset.}
Our experiments are conducted on three popular recommendation system datasets: MovieLens\footnote{https://github.com/wangyu-ustc/DeCA}~\cite{SGDL, DeCA}, Modcloth\footnote{https://github.com/MengtingWan/marketBias}, and Adressa\footnote{https://github.com/WenjieWWJ/DenoisingRec\label{web}}~\cite{SGDL, T-CE, DeCA}. Detailed statistics of these datasets are shown in Table~\ref{tab:dataset statistics}. 
We split each dataset into a training set and a test set in an 8:2 ratio. For ModCloth and Adressa, we follow~\cite{DeCA, T-CE, DCF} to construct the clean test set.

\noindent \textbf{Evaluation Protocols.}
% 现在的一半
Following previous research on denoising recommendation~\cite{SGDL, T-CE, DeCA}, we used three popular metrics: Precision@\textit{N}, Recall@\textit{N}, and MRR@\textit{N}. Higher scores indicate better performance. We conducted each experiment five times and presented the averaged results. 

\vspace{5pt}
\noindent \textbf{Compared Models.}
We compare AEL with several popular top-N recommendation models based on implicit feedback, including GMF~\cite{GMF+NeuMF}, NeuMF~\cite{GMF+NeuMF}, CDAE~\cite{CDAE}, LightGCN~\cite{lightgcn}, TCE and RCE~\cite{T-CE}, and DeCA~\cite{DeCA}.

\vspace{5pt}
\noindent \textbf{Implementation Details.}
The hidden dimensions of sub-AEs are set to 12, 48, and 128, respectively. The number of experts \( k \) activated per input is set to 2. The coefficients of $L_{importance}$ and $L_{Load}$, $w_{i}$ and $w_{l}$, are set to 1e-2. For baselines, we follow the hyperparameter settings suggested in ~\cite{DeCA, T-CE}.

\subsection{Overall Performance Comparison}
In this section, we compare AEL with several prominent top-N recommendation models. Table~\ref{overrall compa} shows the performance comparison on three datasets. The experiment results show that our proposed model outperforms baselines across all datasets. AEL demonstrates good stability and performance across three different datasets, which other models do not possess. We also observe that our model maintains stable performance even on sparse datasets. AEL shows significant advantages in handling dynamic noise, while others' performance can fluctuate with great amplitude across datasets. 

\vspace{-0.1cm}
\subsection{Extensive Effectiveness Analysis}
% \vspace{-0.2cm}
\noindent \textbf{Ablation Study.}
\begin{figure}[t!]
    \vspace{-5mm}
    \centering
    % Create a container to hold the subfigures
    \begin{minipage}{0.23\textwidth}
        \centering
        \includegraphics[width=\textwidth]{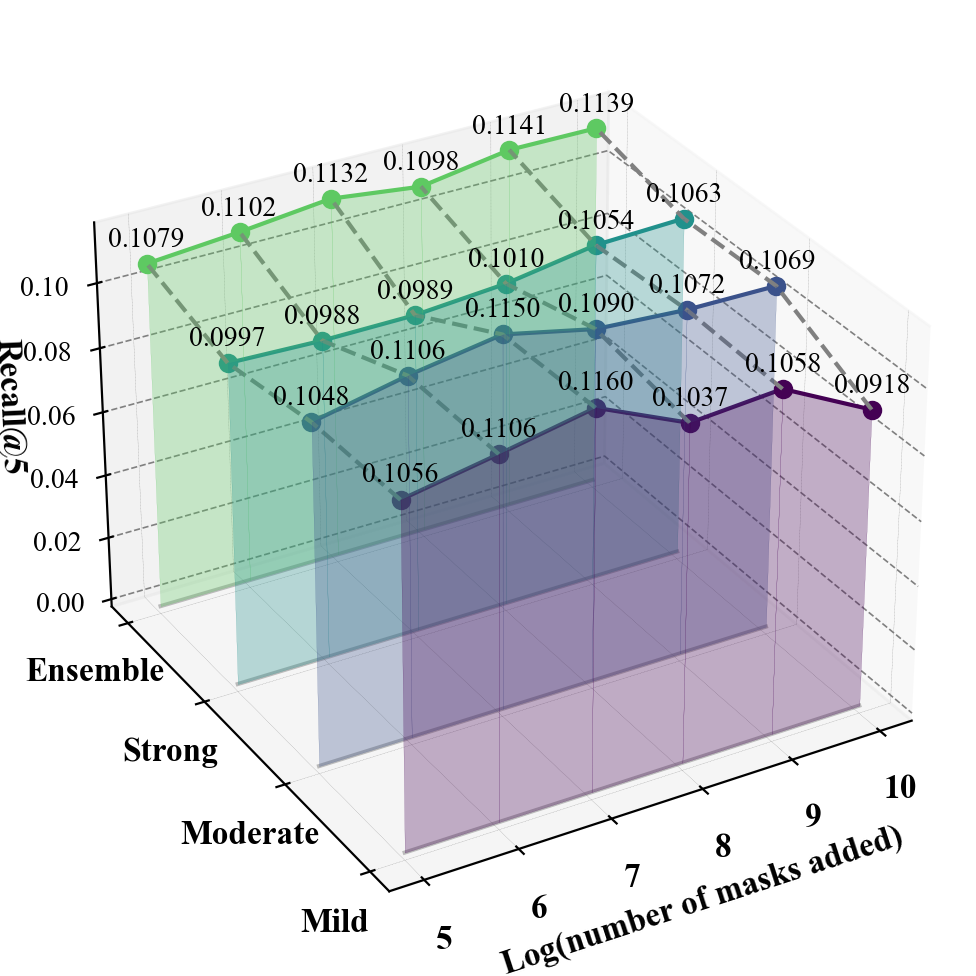}
        % \caption*{(a) Recall@5}  % Caption without a number
    \end{minipage}
    \hfill
    \begin{minipage}{0.23\textwidth}
        \centering
        \includegraphics[width=\textwidth]{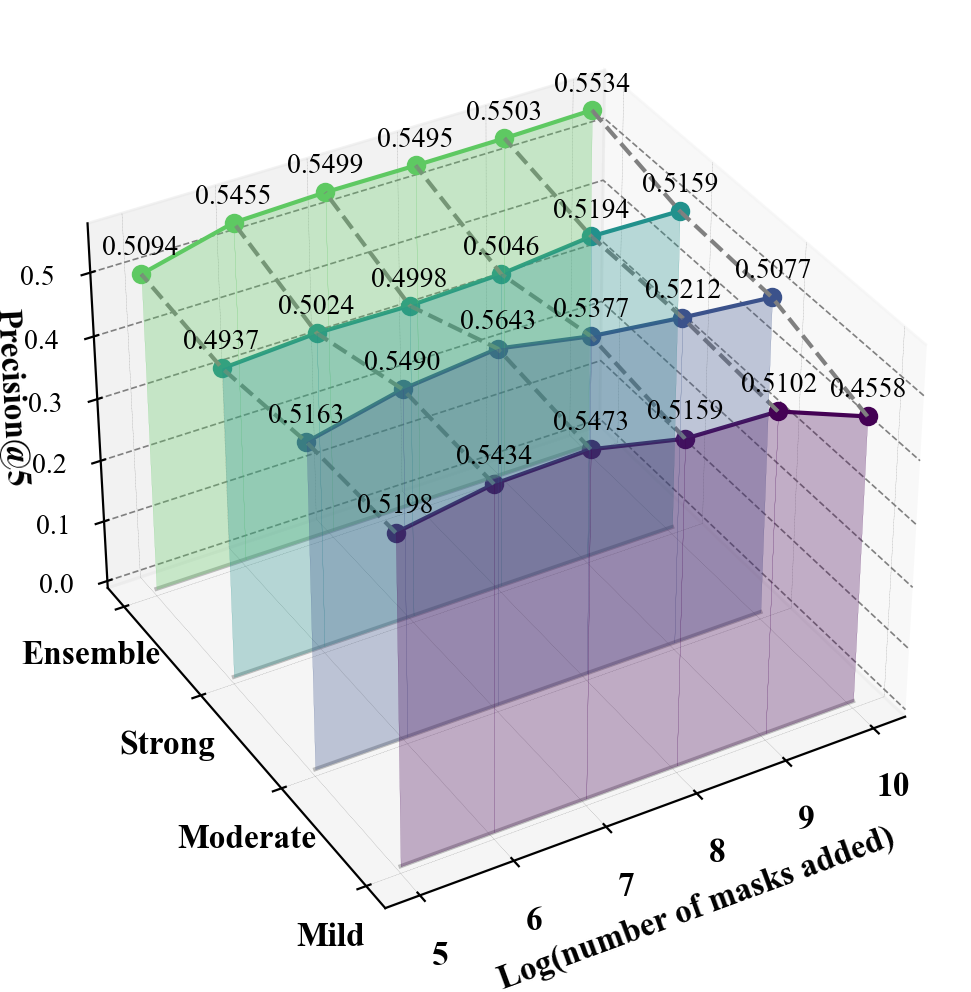}
        % \caption*{(b) Precision@5}  % Caption without a number
    \end{minipage}
    \vspace{-0.1cm}
    \caption{Relationship between performance and noise intensity.}
    \label{fig:trend}
    \vspace{-0.4cm}
    \vspace{-2pt}
\end{figure}
To validate the contributions of parent-AEs, we gradually corrupted the training data by adding noise. To exclude the impact of the sparse gating network on results, we replaced it with averaging the output of all parent-AEs. %Due to the page limitations, 
We present the comparison on the MovieLens dataset in Fig.~\ref{fig:trend}.
The experimental results show that each parent-AE has its own ``comfort zone": Mild Denoising outperforms the others when the number of masks added is small but is surpassed by Moderate Denoising as the number increases. Subsequently, Strong Denoising surpasses both of them. The ensemble model effectively leverages the advantages of all parent-AEs, achieving superior performance and stability.

\noindent \textbf{Hyperparametric Sensitivity Analysis.} The number of experts \( k \) selected is the most critical hyperparameter, as it controls how to synthesize the denoising capacity for current input. We test all feasible \( k \) values, as shown in Table~\ref{k comparison}. We observed that AEL performs well across three datasets in all cases of \( k \), with particularly strong performance when \( k =2\). That reason is that for each interaction vector $x_u$, activating two suitable parent-AEs offers sufficient denoising diversity while avoiding unnecessary experts. Setting \( k=2 \) helps prevent sending inappropriate samples to a parent-AE that inadequately or overly denoise them.

\begin{figure}[t]
\vspace{4pt}
    \centering
    \begin{minipage}{0.23\textwidth}
        \centering
        \includegraphics[width=\textwidth]{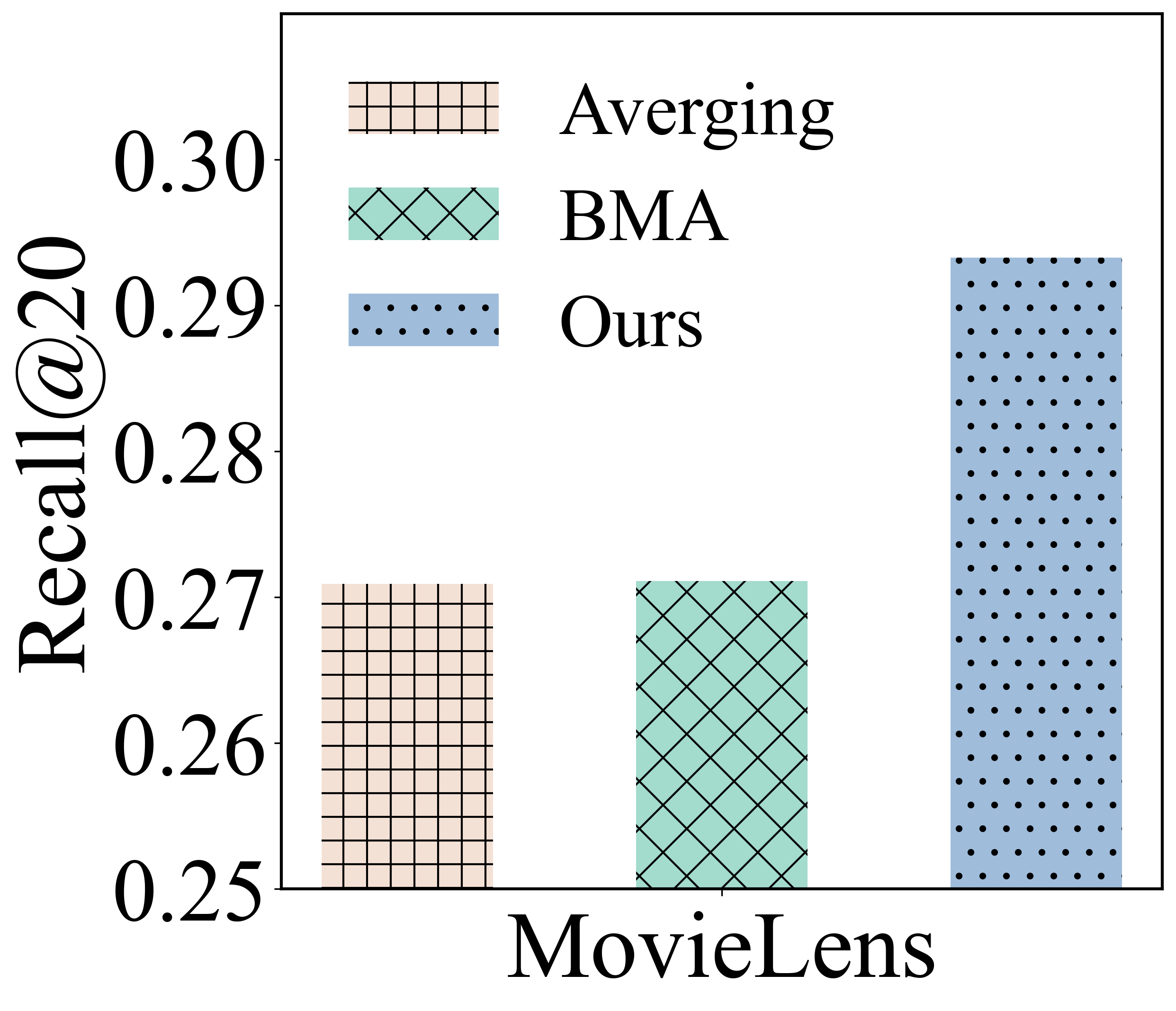}
        % \caption*{(a) MovieLens}  % 使用 caption* 来避免自动编号
    \end{minipage}
    \hfill
    \begin{minipage}{0.23\textwidth}
        \centering
        \includegraphics[width=\textwidth]{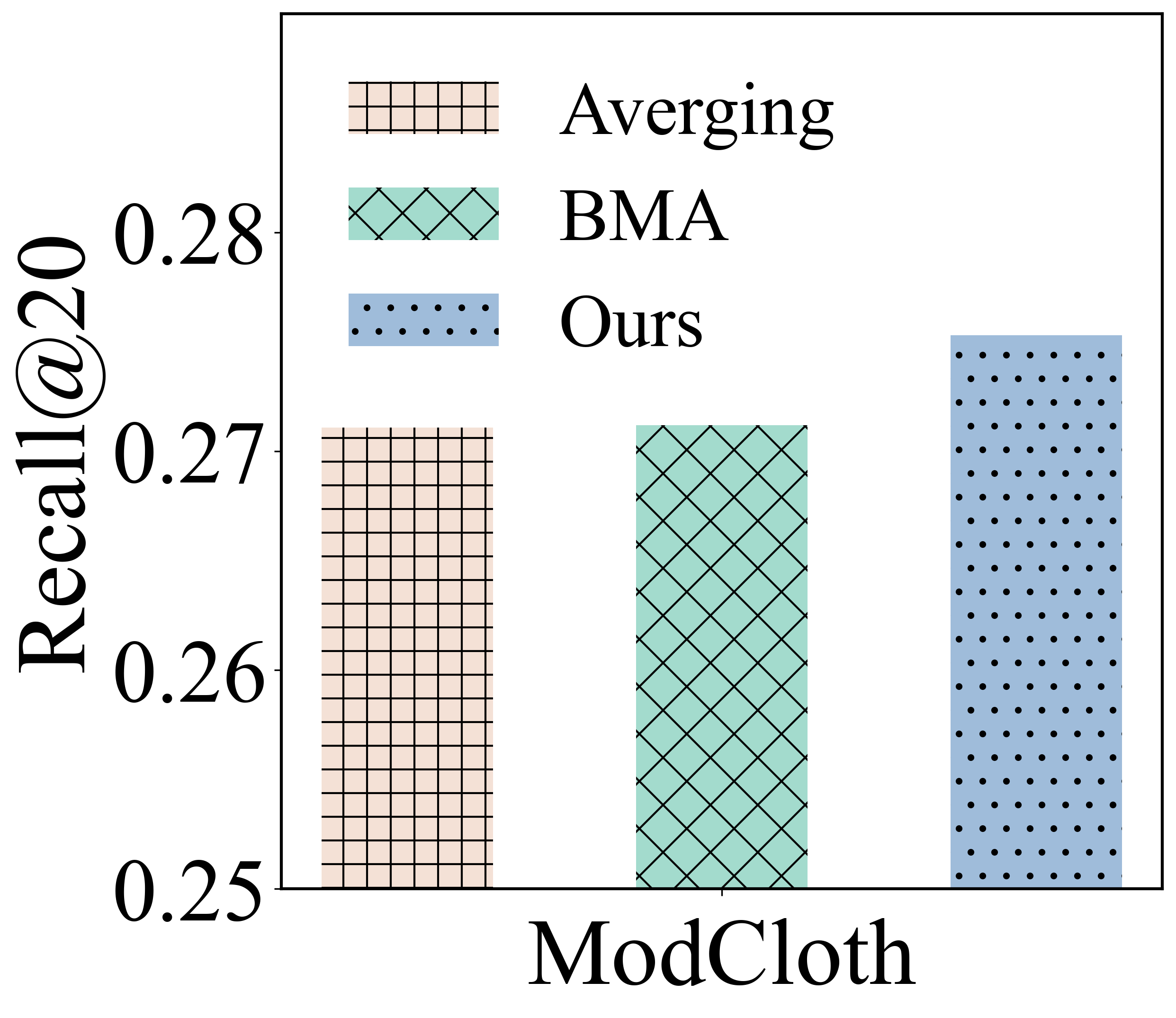}
        % \caption*{(b) ModCloth}  % 使用 caption* 来避免自动编号
    \end{minipage}
    \vspace{-0.1cm}
    \caption{Performance comparison of aggregation methods.}
    \label{fig:agg_comp}
    \vspace{-0.5cm}
    \vspace{2pt}
\end{figure}

\noindent \textbf{Aggregation Method Comparison.}
We conduct experiments to explore how the aggregation methods affect AEL performance. We compare other ensemble learning methods, including Bayesian Model Averaging (BMA) and Averaging, to our spare gating network. According to the results shown in Fig.~\ref{fig:agg_comp}, our aggregation method outperforms others across all datasets, especially on the MovieLens dataset.

\section{CONCLUSION}
In this paper, we proposed \textbf{\uline{A}}daptive \textbf{\uline{E}}nsemble \textbf{\uline{L}}earning (AEL), a novel ensemble method that can automatically synthesize appropriate denoising capacity for different implicit feedback. It contains three parent-AEs as experts and a sparse gating network as a brain. We creatively stack sub-AEs to construct parent-AEs instead of independently creating them, which can effectively reduce model size and ensure denoising capacity diversity. Experiments demonstrate the effectiveness and generalization of our proposed method.

\newpage
\renewcommand{\baselinestretch}{1.1}
\bibliographystyle{IEEEtran}
\bibliography{acml24}
% \bibliography{strings,refs}
\end{document}